\documentclass[12pt,psf,epsf]{JHEP3}
\usepackage[centertags]{amsmath}
\usepackage{graphicx}
\usepackage{epsfig}
\usepackage[mathscr]{eucal}
\usepackage{multicol}
\usepackage{graphicx}
\usepackage{amssymb,latexsym,amsmath}
\usepackage{mathrsfs}
\usepackage{tgothic}

\newcommand{\be}{\begin{equation}}
\newcommand{\ee}{\end{equation}}
\newcommand{\bea}{\begin{eqnarray}}
\newcommand{\eea}{\end{eqnarray}}

\newcommand{\lp}{\left}
\newcommand{\rp}{\right}

\title{Colliding Hadrons as Cosmic Membranes and Possible Signatures of Lost Momentum}
\author{I.Ya. Aref'eva \\
 \small{Steklov Mathematical Institute, Russian Academy of Sciences,}\\
\small{Gubkin str. 8, 119991, Moscow, Russia}
}

\abstract{

We argue that in the TeV-gravity scenario high energy hadrons colliding on
the 3-brane embedded in  $D=4+n$-dimensional spacetime,  with $n$ dimensions
smaller than the hadron size,  can be considered as  cosmic membranes.
In the 5-dimensional case these cosmic membranes produce effects similar to cosmic strings in the 4-dimensional world. We calculate the corrections to the eikonal approximation for the gravitational scattering of partons due to   the presence of effective hadron cosmic  membranes.Cosmic membranes dominate  the momentum lost in the
longitudinal direction for colliding particles that opens new channels for particle decays.
}


\begin{document}

\section{Introduction}

In recent years
the study of transplanckian scattering\footnote{Scattering at
center-of-mass (CM) energies exceeding the quantum gravity scale.}
within the TeV-gravity scenario \cite{ADD} has attracted
significant theoretical and phenomenological interest.
Within the TeV-gravity scenario \cite{ADD} transplanckian scattering
could be observed at the LHC and other future colliders
\cite{factory,GRW98,GRW01,Lonnblad,Boelaert:2009jm,Sessolo:2008ex,SRychkov},
as well as in collisions of high-energy cosmic
neutrinos with atmospheric nucleons
\cite{BHinCR,EMR01}.

Different
physical pictures are expected for different  ranges of impact
parameters $b$. For  impact parameters $b$ of the order of the
Schwarzschild radius $R_{S}$ of a black hole of mass $\sqrt{s}$,
microscopic black hole formation and its subsequent evaporation is
expected \cite{BF99,IA99,EG02,GR04} \footnote{See also \cite{IA09} and refs.
therein; there are also proposals concerning the production of more
complicated objects such as wormholes, or time machines
 \cite{AV-TM,MMT,NS}.}, while for large impact parameters $b\gg R_{S}$
 the eikonal picture
 given by eikonalized
single-graviton exchange is expected \cite{tHooft,ACV87,Verlindes,Kabat92}.
Corrections in $R_{S}/b$ to the elastic eikonal scattering have been studied
\cite{Amati:1987uf,Amati:1992zb,areva,Ciafaloni:2008dg}.

To study high-energy scattering of the hadrons
one usually deals with the parton picture. In the case of a 3-brane embedded in
$D=4+n$-dimensional spacetime for  large impact parameters
graviton exchanges dominate in  parton amplitudes
\cite{tHooft,Verlindes,Amati:1987uf,Kabat92,GRW98}.
In all $D$  dimensions the graviton is supposed to be propagated
freely. Since $D$-dimensional gravity is strong  it would be interesting
to calculate the modification of the graviton
propagator due to a presence of matter. This is difficult problem, however
it can be solved  in 2+1 gravity, where we know
analytically
the modification of the spacetime due to the presence of
pointlike matter. We know also the modification of the spacetime metric
by a cosmic string in  4-dimensional spacetime and by a cosmic membrane
in 5-dimensional spacetime.

Due to Lorentz contraction we can treat  colliding
hadrons in the laboratory frame  as  membranes with the transversal characteristic
scale of order of the  hadron and a negligible thickness.
 These membranes are located  on our 3-brane.  Since $4+n$ gravity
 is strong enough we can expect that hadron membranes
 modify  the $4+n$-spacetime metric.

 Only for the case of $n=1$ we know explicitly the
 modified metric and we can estimate explicitly the influence of this
  modification on the
 parton and other particle scattering. It is known that the
 5-dimensional ADD model with $M_{Pl,5}\sim $ TeV
 is not phenomenologically acceptable and we
 can  deal with the RS2 model \cite{RS2} or with the DGP model \cite{DGP}.
In all these  cases we treat a moving hadron
as an infinite moving membrane  in the 5-dimensional world with location
on the 3-brane (our world). In other words, we deal with an effective
3-dimensional  picture
in the high-energy scattering
(compare with the usual effective  2-dimensional picture in 4-dimensional spacetime,
see \cite{IA,LL} and refs therein).

In the framework of the  picture described above,
we can consider the influence of the
matter on graviton propagation. Due to the presence of the
hadron membrane the gravitational background is nontrivial and
describes a flat spacetime with a conical singularity  located on
the hadron membrane. This picture is a generalization of the
cosmological string picture in the 4-dimensional world to  the
5-dimensional world. The deficit angle  is  proportional to the product
of the hadron matter density on the membrane and the 5-dimensional
gravitational coupling. This is a rather small number \footnote{One can compare this number with an
estimate of the deficit angle  $\delta _{cs}\sim 10^{-6}$ for a cosmic string in
 4-dimensional spacetime
with the Newtonian gravitational constant $G_{N,4}$
and the density
$\rho=\frac{m}{l}=10^{33}GeV^2$, that corresponds  to the Earth mass
distributed
 on a length of  about $l=9$ km.
}, $\delta_{h_0}\sim \frac{1}{M_{Pl,5}^3}
\frac{M_{hadron}}{l^2_{hadron}}\sim 10^{-9}$. Since the hadrons collide with
Lorentz boost factor, $\gamma=1/\sqrt{1-v^2}$, about $\gamma\sim 10^{4}$,  we have
$\delta_{h}\sim 10^{-5}$. For heavy ions  composed of $A$
hadrons,
this number is near
$\delta_{Ion}\sim A_{Ion}^{1/3}\delta_h$.

We can take into account
corrections to the graviton propagation. A study of these corrections
and their physical consequences is the subject of the present letter.
A more detailed discussion of
the topological defects in  TeV-gravity including the RS2 and DGP models
and   will be presented in
\cite{IA-AB}.
 As to higher dimensional cases we can just expect that numerical calculations
could exhibit similar qualitative results.

The paper is organized as follows. In Section 2  we present our setup and argue
 why in the TeV-gravity scenario the high energy hadrons colliding on
 the 3-brane embedded in $4+n$-dimensional spacetime  with $n$ dimensions
smaller than the  hadrons size,  can be considered as  cosmic membranes
 in the $4+n$-dimensional world. We recall basic
 facts about
eikonalization of graviton exchanges and the form
of the spacetime metric with a cosmic membrane. In Section 3 we
present corrections to
 the eikonal
phase due to a conical singularity.
We restrict ourself here  to a flat bulk  for simplicity.
The AdS case corresponding to the RS2 model can be investigated in a similar way.
  Others possible effects  related with cosmic membranes  and their signatures
are briefly discussed in the conclusion.

\section{Setup}
It is known that  for large impact parameters $b\gg R_{S}$
(elastic small-angle scattering) the transplanckian  amplitude
is dominated by eikonalized single-graviton exchange
\cite{tHooft},\cite{ACV87},\cite{Verlindes},\cite{Kabat92}.
The eikonal amplitude has been used in \cite{EMR01} to compute the
differential cross section for neutrino-nucleon scattering and in \cite{GRW01}
to compute the close to beam jet-jet production  at the LHC.
For small impact parameters $b\ll R_{S}$ the nonlinear effects are important and
within the classical gravity one can expect the black hole formation.

\subsection{Hadron as a membrane in 5-dimensional world}
The graviton exchange is supposed to take place in the $4+n$-dimensional spacetime.
In the total transplanckian cross section  there is a factor, describing dependence
on $n$ and on the form of the background in the extra dimensional spacetime.
In all previous considerations \cite{EMR01,GRW01,SRychkov}  the graviton is supposed
to  propagate
freely in  extra dimensions. It would be interesting to be able to calculate the
modification of the propagator due to the presence of the hadron matter.
This can be done for example in the 2+1 gravity, where we know
analytically the modification of the spacetime due to the present of pointlike matter.

 In $2+1$ dimensions, solutions to Einstein's equation
with  point masses are flat metrics except  conical singularities at the
location of the masses. In $3+1$ dimensions,
there are solutions with singularities on the worldsheets of the strings.
The deficit angle of the conical singularity
is proportional to the mass in the $2+1$ case and the mass per length $\mu$ in the $3+1$ case
\cite{DJH84}. In $4+1$ dimensions,
there is a solution with singularity on the worldsheet of
the membrane. One can  imagine this membrane as  high velocity moving hadron,
that in the rest
frame is tried as a ball. If we have  extra dimensions, they are not available
for the hadron
and  the hadron membrane cannot stretch in these dimensions.
Hence,  we get the 2-dimensional hadron membrane propagated on the 3-brane embedded
in $4+n$-dimensional spacetime.  We know  explicit  solutions
to Einstein's equations with the hadron membrane
in the 5-dimensional ADD
 and  RS2 models. The first case is simpler and in spite of it is not phenomenologically
 acceptable,  we consider this case for simplicity \footnote{One can assume
 an anisotropic compactification with essentially
suppressed $n-1$ dimensions (in this case $M_{Pl,D}\sim $ TeV and $M_{Pl,5}\sim 10^3$ TeV),
or just consider a toy model with $M_{Pl,5\,toy}\sim 10^3$ TeV.}.

 \begin{figure}[h!] \centering
\includegraphics[width=5cm]{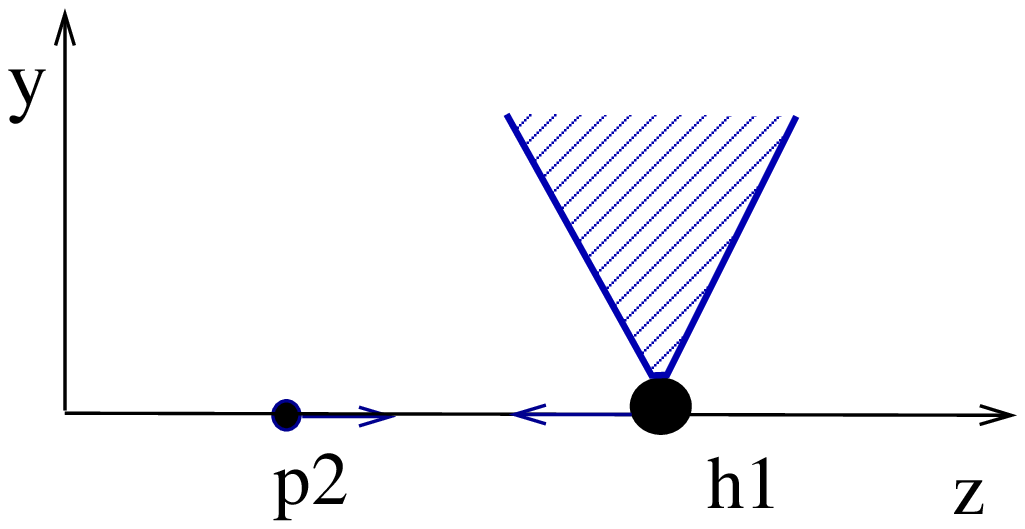}$\,\,\,\,$A.
$\,\,\,\,\,\,\,\,\,\,$
\includegraphics[width=5cm]{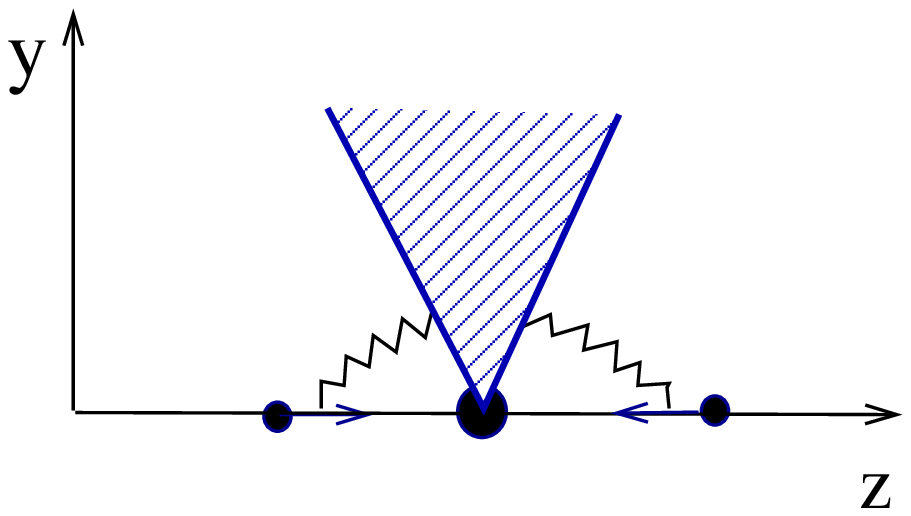}$\,\,\,\,$B.
\label{Fig:hex} \caption{A. Ultra
relativistic colliding hadron $h_1$ as it  seen by the parton $p_2
$. B. The graviton exchange with modified propagator
between partons. }
 \end{figure}

\subsection{Bulk with conical singularities}
In the ADD model the metric in the bulk is flat,
\begin{equation}
\label{M5} ds^2= -dt^2 + d\vec{x}_\perp^2 + d\rho^2 + \rho^2 d\Omega^2 ,\,\,\,\rho^2=
\sum _1^n y_i^2+z^2,\,\,\,\vec{x}_\perp=(x_1,x_2)
\end{equation}
here $x_1,x_2,y_i, z$ are coordinates in the bulk and $d\Omega^2$ is the metric
on the unit sphere $S^{n}$.
 However, the hadron membrane
produces a nontrivial background. We know this background explicitly
for the case of $n=1$. In this case the bulk metric remains locally
flat, $d\Omega^2=d\phi^2$ and the hadron membrane produces only the
conical singularity, i.e. the range of the angle is $0<\phi<\alpha$.
The angle $\alpha$ defines the deficit angle  $\delta$
\be
\alpha=2\pi-\delta,
\ee
where
\be \label{dif}
\delta=8\pi G_5\frac{m_h}{S_h}
=\frac{32}{M^3_{Pl,5}} \frac{m_h}{ l^2_h}.
\ee
 here $m_h$ is the hadron mass and $l_h$ is the hadron size, $S_h=\pi l^2_h/4 $. The top of the cone is located on the brane.

 The gravitational
effect of the hadron membrane in the RS2 model  is convenient to present in the
Poincar\'e coordinates. Starting from the metric \be
\label{RS2}
ds^2=a^2(y)\eta _{\mu\nu}dx^\mu dx^\nu+dy^2,\ee
where $\eta _{\mu\nu}$ is the 4-dimensional Minkowski metric and the warp factor
$a(z)$ has the form \cite{RS2}
\be
a(y)=e^{-k|y|},\ee
$1/k$ is the radius of 5-dimensional AdS spacetime,
 we get the metric in the Poincar\'e coordinates after the following change of variable,
 $y \,\to \,w$,
$ w=r_0e^{y/r_0}$,
 \be
ds^2=\frac{r_0^2}{w^2}(\eta _{\mu\nu}dx^\mu dx^\nu+dw^2)\ee
\cite{vilenk85}. According to the usual prescription to incorporate a
 membrane
we cut a wedge.
This
 can  be done by reducing the range of a suitable  angular
 coordinate. For example, for $AdS_5$
\begin{equation}
\label{As-CS-naive} {R_5^2\over
w^2}\lp[dw^2 -dt^2 + dz^2 + d\rho^2 + \rho^2 d\phi^2 \rp]~,
\end{equation} %
and the range of the angle is $0<\phi<2\pi-\delta$ where $\delta$ is given by (\ref{dif}).

\subsection{Eikonalization of graviton exchanges}
The parton-parton elastic forward scattering amplitude
for a large center of mass energy is given by the eikonal  technique
\cite{Eikonal},\cite{Eikonal-dubna}. In the transplanckian regime
 the graviton exchanges \cite{Kabat92,GRW01} dominate and define the amplitude
\begin{equation}
\mathcal{A}_{\rm eik}(\mathbf{q})=\mathcal{A}_{\rm Born}+\mathcal{A}
_{1-loop}+\ldots=-2is \int d^{2}\mathbf{b}\,e^{-i\mathbf{q}.\mathbf{b}
}(e^{i\chi(\mathbf{q})}-1)\,, \label{eq:AeikR}
\end{equation}
where the eikonal phase $\chi$ is given by the Fourier transform of the
Born amplitude in the transverse plane
\be
\label{phase-flat}
\chi(\mathbf{b})=\frac{1}{2s}\int\frac{d^{2}\mathbf{q}}{(2\pi)^{2}%
}e^{i\mathbf{q}.\mathbf{b}}A_{\text{Born}}(s,\mathbf{q})\,.
\ee
 The
 $4+n$-dimensional
 Born amplitude for the  exchange of the graviton, which does not get any
 transferred momenta in the  direction
 transversal to the brane, is given by
 \be
 \label{Born-extra-dim}
 \mathcal{A}_{Born}(s,q)=
 \frac{-s^2 }{ M_D^{n+2}} \int  \frac{d^nl}{ \mathbf{q}^2 +l^2},\,\,|\mathbf{q}|=q.
 \ee

\begin{figure}[h!]
$$\,$$
\begin{center}
\includegraphics[width=3cm]{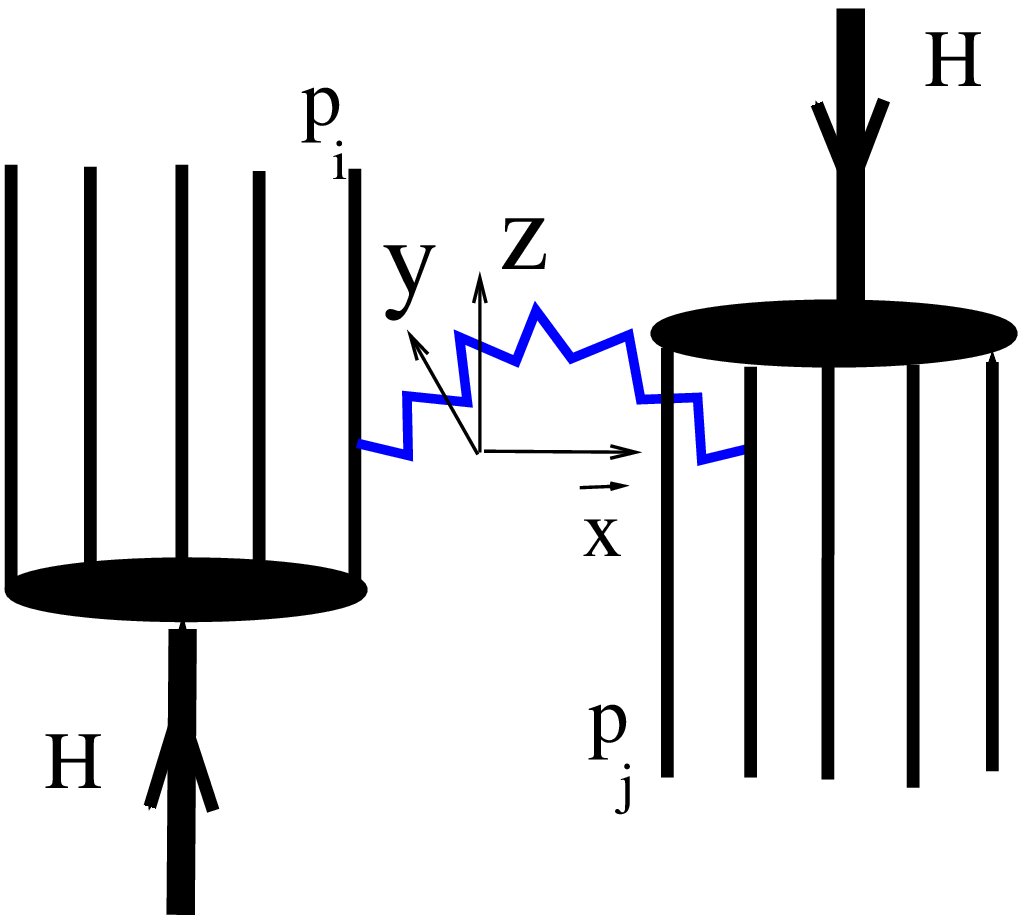}$A\,\,\,\,\,\,\,\,\,\,\,\,\,\,\,\,\,\,$
\includegraphics[width=7.5cm]{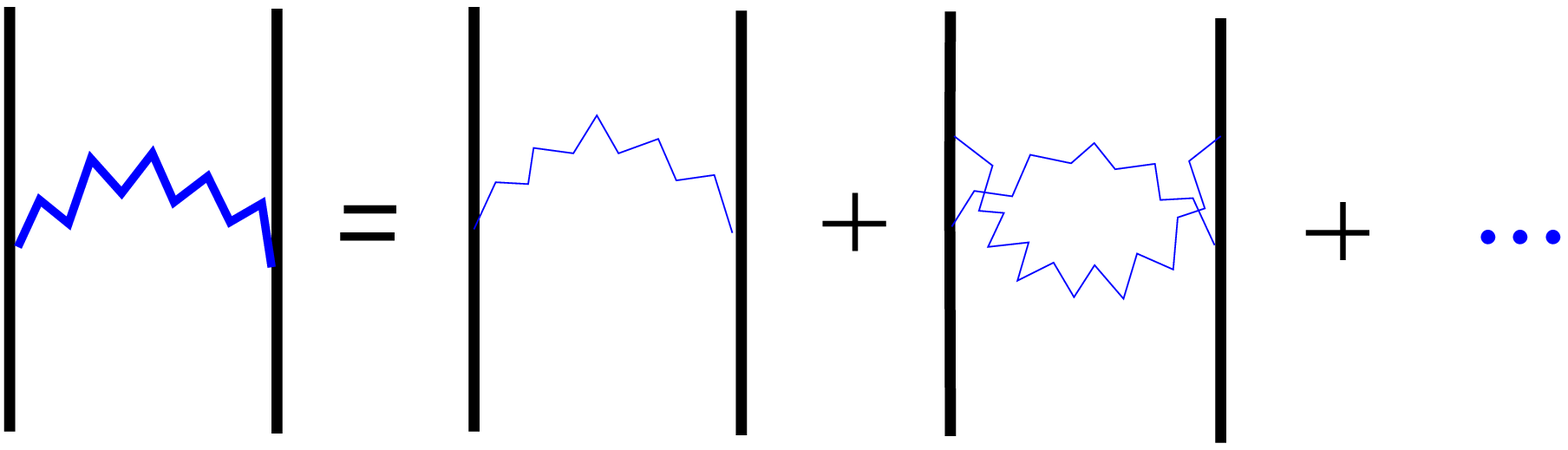}$B$
\end{center}
\caption{ A. Collision of hardons with a large impact
  parameter in $(x_\bot,z)$ coordinates  is presented as an elastic scattering between
  partons due to a free graviton exchange. $y$-coordinate schematically presents extra
  dimensions.
  B. The $2\to2$ small angle transplanckian scattering amplitude is given
by a sum of crossed-ladder graviton exchanges.} \label{Fig:DISpp}
\end{figure}

The expression
for the eikonal amplitude~\cite{EMR01,GRW01} is
\be
{\cal A}_{\rm eik} =4\pi s b_c^2 F_n (b_c q),
\label{eikfin}
\ee
\be
F_n (y)=-i\int_0^\infty dx x J_0 (xy) \left( e^{ix^{-n}}-1\right) ,
\label{funn}
\ee
where the integration variable is  related with the impact parameter, $x=b/b_c$
and in (\ref{funn}) we take into account that the  eikonal phase has
the power dependence on the impact parameter
  \be
\chi(b) =\left( \frac{b_c}{b}\right)^n, \,\,\,\mbox{where}\,\,\,b_c\equiv \left[ \frac{(4\pi)^{\frac{n}{2}-1}s\Gamma (n/2)}{2M_D^{n+2}}
\right]^{1/n}.
\label{bbc}
\ee

Functions $F_n$, $n>1$, when  $y\gg 1$ oscillate around their asymptotic values
given by $F_{n,as}(y)=\frac{-in^{\frac{1}{n+1}}y^{-\frac{n+2}{n+1}}}
{\sqrt{n+1}} \exp \left[ -i(n+1)\left( \frac{y}{n}\right)^{\frac{n}{n+1}}
\right] $ \cite{GRW01}.  Within the TeV-gravity scenario \cite{ADD} the total
transplanckian cross section
 is
finite, grows with energy, and is dominated by
small-angle scattering between partonic constituents
\cite{EMR01},\cite{GRW01}.

The real and imaginary parts of the function $F_1$ are shown in Fig.~3.A,
and we also see the oscillations of the real part of the function $F_1$.
\begin{figure}[h!] \centering
\includegraphics[width=4cm]{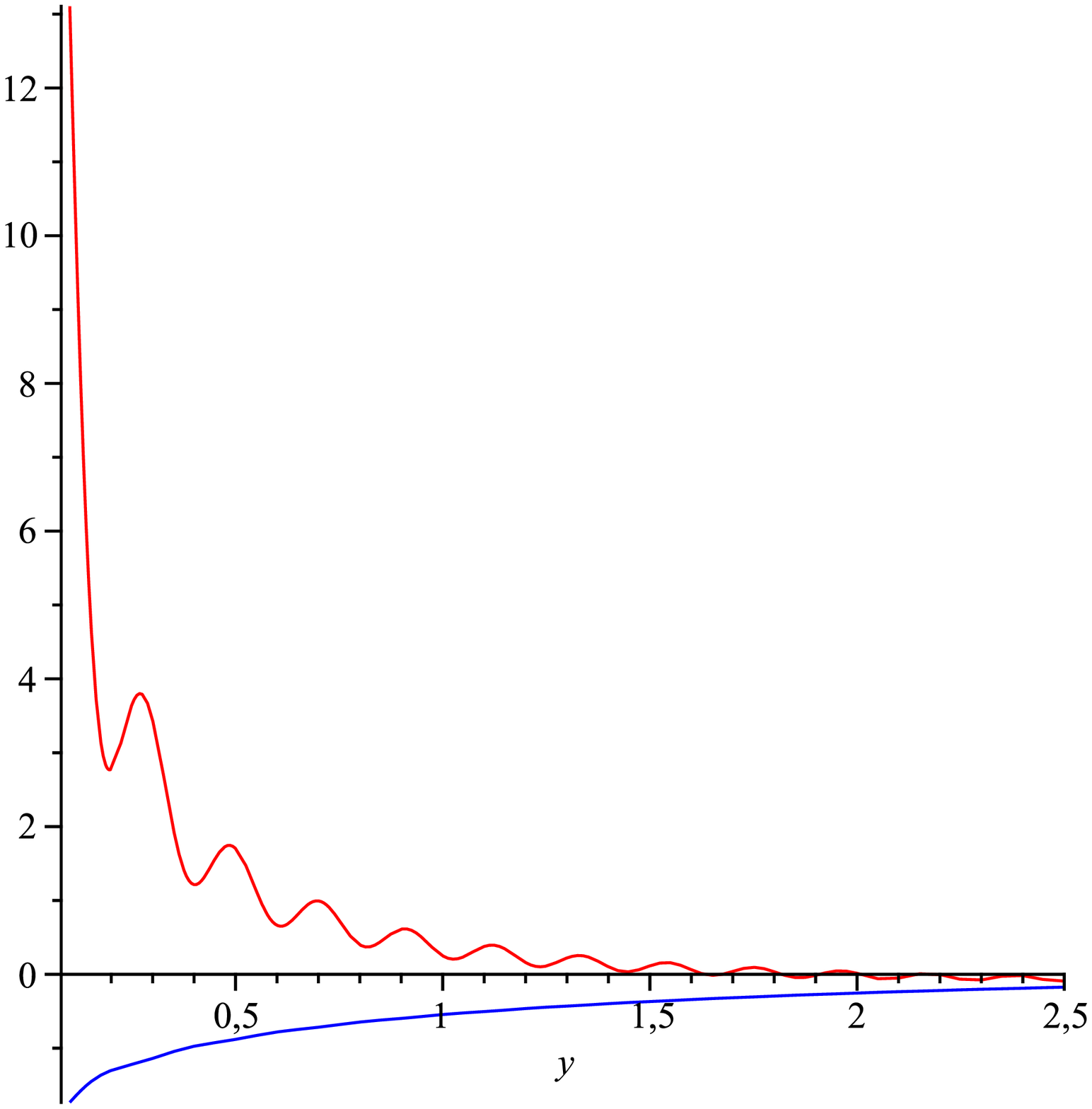}$\,\,\,\,\,$A$\,\,\,\,\,\,\,\,\,\,\,\,$
\includegraphics[width=4cm]{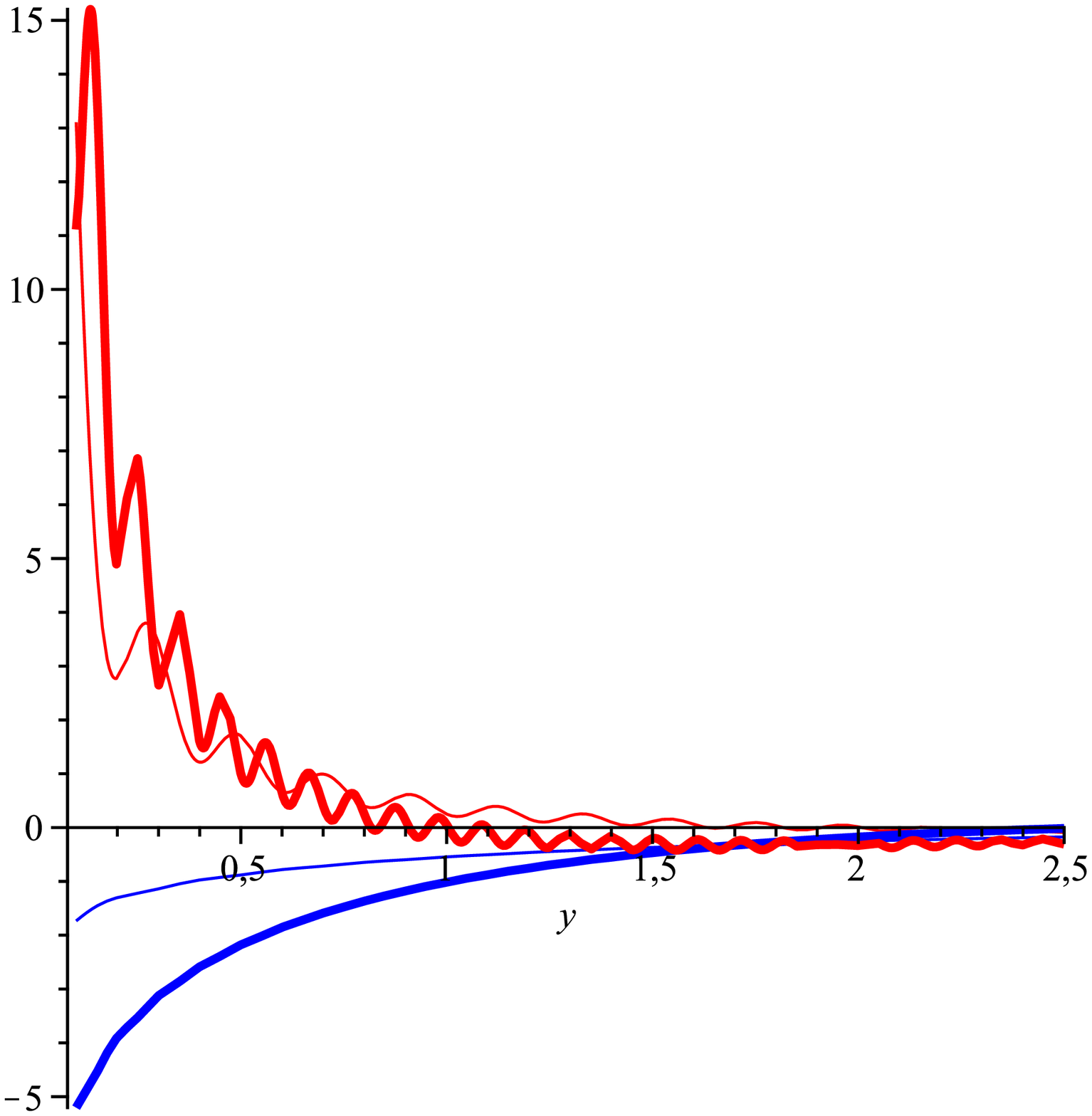}$\,\,\,\,\,$B
\label{Fig:F1} \caption{A. The real (red) and imaginary (blue) parts of the eikonal  amplitude $F_1$.
B. Thick lines represent the real and imaginary parts of the eikonal amplitude
with a doubling eikonal phase in the toy model with the deficit angle  equal to $\pi$.
 }
 \end{figure}

\section{Eikonal in the conical spacetime}
The goal of this section is to estimate the influence of the hadron membrane on the
forward scattering of the partons.

\subsection{Graviton exchange with modified graviton propagator}
The  tree level 2 partons $\to$ 2 partons S-matrix element
corresponding to  one graviton exchange
in the   $C_\alpha \times M^3$ spacetime,
\be
 <p_1,p_2|S|p_3,p_4>_{\rm  1 graviton}\equiv
 {\cal S}_{{\rm   graviton,}\alpha} (p_1,p_2,p_3,p_4),
 \ee
 is given by the linearization of gravity \cite{Kabat92} and in $s\gg t$
regime is
 \be
  {\cal S}_{{\rm   graviton,}\alpha} (p_1,p_2,p_3,p_4)
\approx -  16 \pi G \gamma(s) \,{\cal S}_{scalar,\alpha },
\label{prop-m}
\ee
here $\approx$ means that we ignore the recoil of the matter field and take the
prefactor $\gamma(s)$ the same as for  the  flat case, $\gamma(s)=((s-2m^2)^2-2m^4)/2$.

In the flat spacetime
\be
{\cal S}_{\rm graviton, flat}(p_1,p_2,p_3,p_4)=i(2\pi)^4\delta^4(p_1+p_2-p_3-p_4)\,
{\cal A}_{\rm Born}(s,t), \,\,\,\,\,t\approx-\mathbf{q}^2.
\ee

\begin{figure}[h!]\centering
\includegraphics[width=5cm]{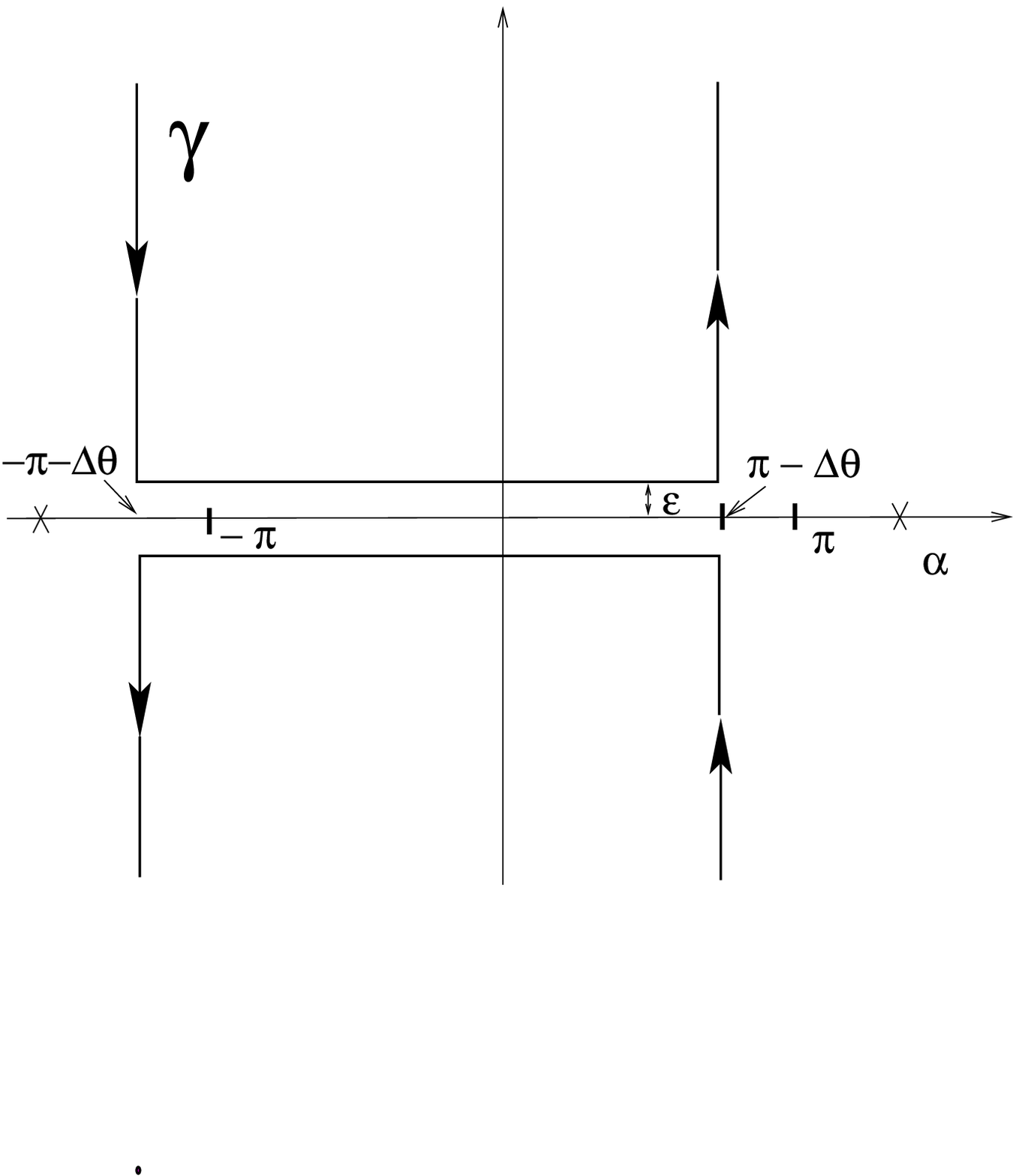}$\,\,\,\,\,\,\,\,\,\,\,\,\,\,$
\includegraphics[width=5cm]{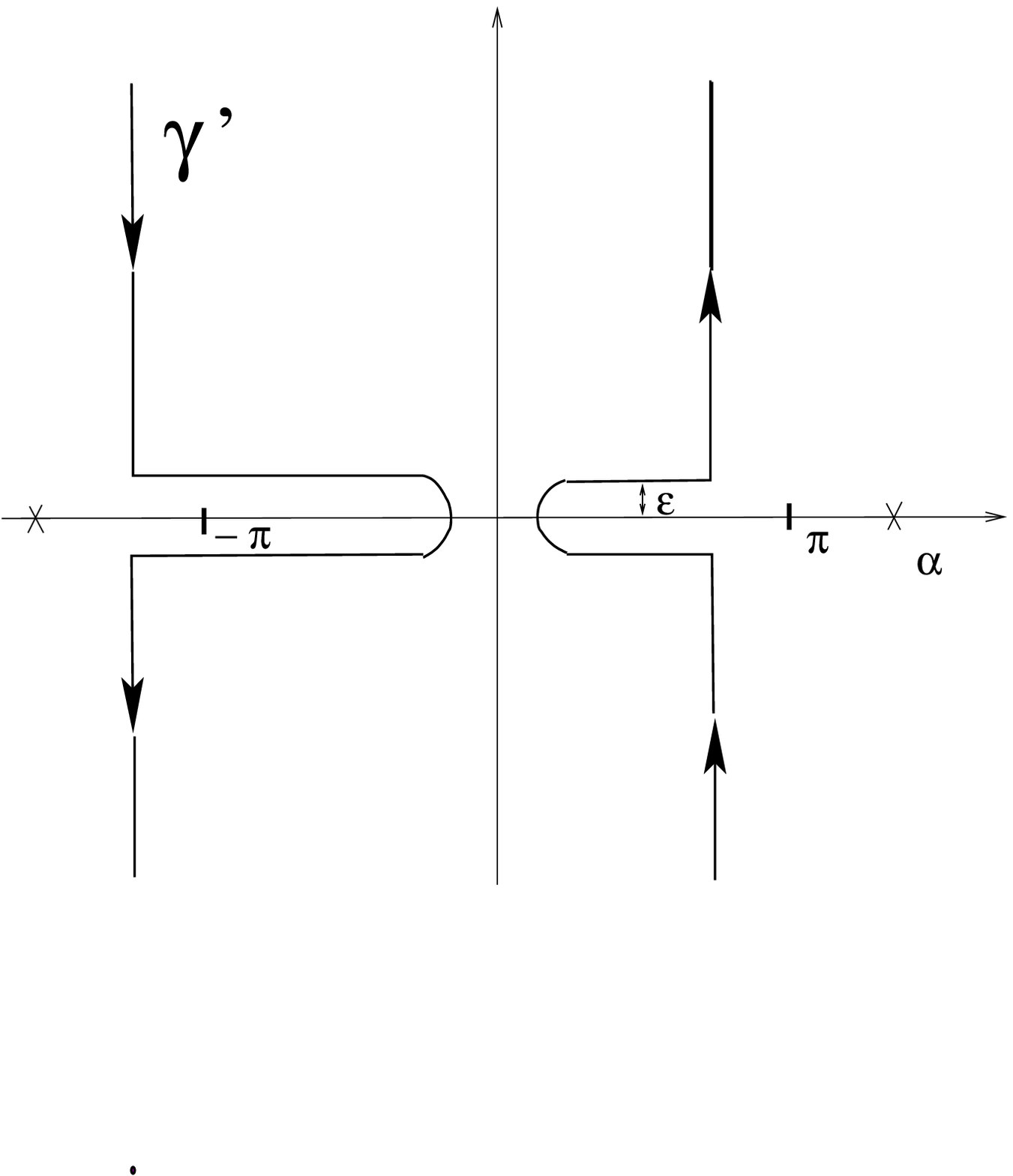}
\caption{ Contours $\gamma$ and
 $\gamma^\prime$.}
\label{Fig:contours}
\end{figure}

In what follows,
${\cal S}_{scalar,\alpha }\equiv {\cal S}_{\alpha }$ is the Born amplitude for
the scalar particles
scattering due to the scalar exchange in  the $C_\alpha \times M^3$ spacetime.
It  can be written  (after the Euclidean rotation) in the Schwinger representation as
\bea
\nonumber
{\cal S}_{\alpha }=\int d^4Xd^4X'e^{i(p_1-p_3)X+i(p_2-p_4)X'}\int
d\tau e^{-m^2\tau}
K(t,x_{\bot }; t',x'_{\bot };\tau)K_{\alpha}(z,0;z',0;\tau),
\eea
here $X=(t,x_{\bot },z)\equiv (x^{\check{\mu}},z)$
and $K(t,x_{\bot }; t',x'_{\bot };\tau)$ is the heat kernel on  the 3-dimensional plane
 and $K_{\alpha}(z,y;z',y';\tau)$ is the heat kernel on the 2-dimensional cone $C_\alpha$.
 $K_{\alpha}$ has a representation \cite{Dowker77,DJ88,Hooft88}
\be
K_{\alpha}(z,y;z',y';\tau)=\frac{i}{2\alpha }\int_\gamma dw\,
{\mbox{ctg}\,}\left(\frac{\pi w}{\alpha}\right)\, K(z(w),y(w);z',y';\tau).
\label{HKreprB}
\ee
Here
$
\left(z(w),y(w)\right)=\left(r\cos(\theta+w), r\sin(\theta+w)\right)$,
$(r, \theta$) are related with coordinates $(z,y)$ as
$\left(z,y\right)=\left(r\cos(\theta), r\sin(\theta)\right)$,
$K(z,y;z',y';\tau)$ is the heat kernel on  the 2-dimensional plane
\be
K(z,y;z',y';\tau)=\frac{1}{4\pi \tau} \exp\{-\frac{(z- z')^2+(y- y')^2}{4\tau}\},
\ee
and $\gamma$ is a characteristic contour presented in Fig. \ref{Fig:contours}, where  $\Delta \theta=
\theta^\prime -\theta$ and $\theta '$ is related with $z'$.

Under assumption that we are on the brane, $\theta=0$ and $\theta'=0$ (or $\theta,\theta'=\pi$) we have
\be
(z(w),y(w))|_{\mbox {on brane}}=\left(z\cos(w), z\sin(w)\right), \,\,\,\,
(z',y')|_{\mbox {on brane}}=(z',0)\ee
and
\be
K_\alpha(z,0;z',0;\tau)=\frac{i}{2\alpha }\int_\gamma dw\,
{\mbox{ctg}\,}\left(\frac{\pi w}{\alpha}\right)\,{\cal K}_{w}(z,z';\tau)
\label{HK-brabe}
\ee
where
\be
\label{HK-r-mm}
{\cal K}_{w}(z,z';\tau)\equiv \frac{1}{4\pi \tau}
\exp\{-\frac{z^2+{z'}^2-2zz' \cos w}{4\tau}\}
\ee
 We can define  the Fourier transformation of the propagator associated with
 (\ref{HK-r-mm}) as
\be
{\cal D}(r,v)=\int \int e^{ir(z-z')+iv(z+z')}e^{-m^2\tau}{\cal K}_{w}(z,z';\tau)
dzdz'\frac{d\tau}{4\pi \tau}
\ee
and find
\be
{\cal D}(r,v)=
\frac{2}{\sin w }\frac{1}{\frac{r^2}{\sin^2\frac{w}{2}}+\frac{v^2}{\cos^2\frac{w}{2}}+m^2}.
\ee

Finally, we get
\bea\nonumber
{\cal S}_{\alpha }&=&i(2\pi)^3\delta^{3}\left((p_1+p_2-p_3-p_4)_{\check{\mu}}\right)
{\cal M}_{\alpha },\\\nonumber\\
{\cal M}_{\alpha }&=&\frac{i}{2\alpha }\int_\gamma dw\,
{\mbox{ctg}\,}\left(\frac{\pi w}{\alpha}\right)
\,\frac{2}{\sin w}\,\frac{1}{\frac{Q^2}{\sin^2 \frac{w}{2}}+
\frac{P^2}{\cos^2 \frac{w}{2}}+q_{\check{\mu} }^2+m^2},
\label{BA-dif1}\eea
here and below $q_{\check{\mu}}=(q_0,q_1,q_2)$, $\check{\mu}=0,1,2,\,\,$
$q=(q_{\check{\mu}},q_z)$, $q_\perp= (q_1,q_2)$,
\be \label{QPqmu}Q=\frac{1}{2}(p_1-p_2-p_3+p_4)_z,\,\,\,
P=\frac{1}{2}(p_1+p_2-p_3-p_4)_z,\,\,\,
q_{\check{\mu} }=(p_1-p_3)_{\check{\mu}}.
\ee
$Q$ and $P$ are related  as
$Q=q_z-P$.
In the eikonal  regime
$Q\approx-P
$
and this gives a simplification of
(\ref{BA-dif1})
\be
{\cal M}_{\alpha }\approx
\frac{i}{2\alpha }\int_\gamma dw\,
{\mbox{ctg}\,}\left(\frac{\pi w}{\alpha}\right){\cal B}_w(q_{\bot},P),\ee
where
\be
\label{Mw}
{\cal B}_w(q_{\bot},P)=\frac{2}{\sin w}\,\frac{1}{q_{\bot}^2+m^2+\frac{4P^2}{\sin ^2 w}
}.\ee

\begin{figure}[h!]
$$\,$$
\begin{center}
\includegraphics[width=3cm]{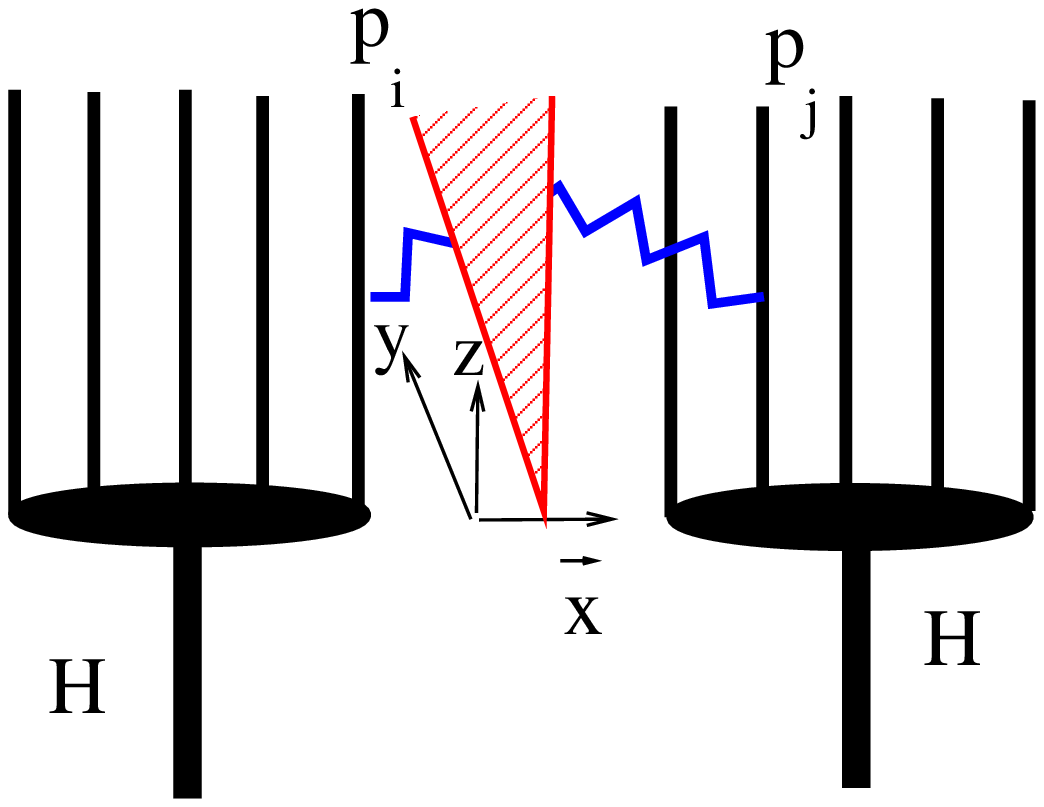}$A\,\,\,\,\,\,\,\,\,\,\,\,\,\,\,\,\,\,$
\includegraphics[width=7.5cm]{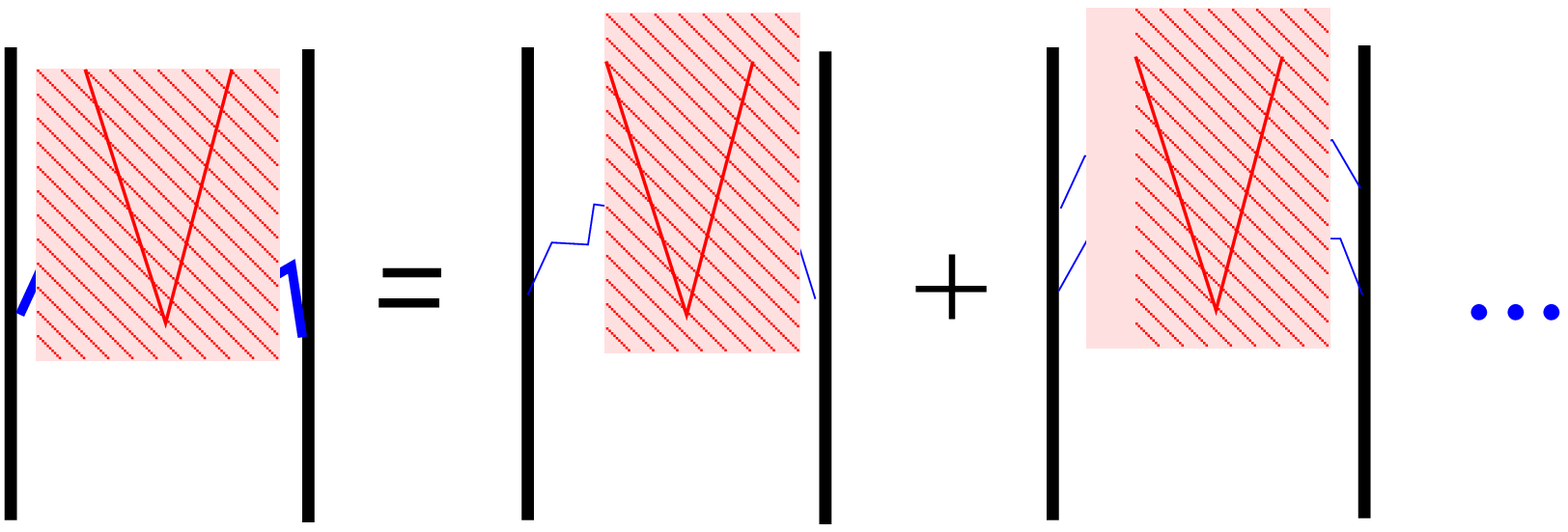}$B$
\end{center}
\caption{ A.Collision of hadrons with a large impact
  parameter   is presented as an elastic scattering between
  partons due to a  graviton exchange in the space $(x_\bot,z,y)$ with the conic
  point in the $(z,y)$ section.
   B. The $2\to2$ small
angle transplanckian scattering amplitude is given by a sum of crossed-ladder
graviton exchanges in the  space with the hadron membrane.} \label{Fig:eikonal}
\end{figure}

Let now define the $w$-eikonal phase $\chi$ as the Fourier transform of (\ref{Mw})
\be
{\cal X}_w(\mathbf{b},P)=\frac{1}{2s}\int\frac{d^{2}\mathbf{q}}{(2\pi)^{2}
}e^{i\mathbf{q}.\mathbf{b}}{\cal B}_w(q_{\bot},P)\,.
\ee
The total eikonal phase is given by the integral over the contour $\gamma$
\be
\chi_\alpha(\mathbf{b},P)=\frac{i}{2\alpha }\int_\gamma dw\,
{\mbox{ctg}\,}\left(\frac{\pi w}{\alpha}\right){\cal X}_w(\mathbf{b},P)\ee
Using the explicit expression for the eikonal phase for a massive particle we get
\be
{\cal X}_w(\mathbf{b},P)=\frac{1}{2\tau}\frac{1}{\pi\sin w}
\,K_0\left(|\mathbf{b}|\sqrt{m^2+\frac{4P^2}{\sin ^2 w}}\right).
\ee

In the case of $m\approx 0$
\be
{\cal X}_w(\mathbf{b},P)=\frac{1}{2\tau}\frac{1}{\pi\sin w}
\,K_0(2|\mathbf{b}||\frac{P}{\sin w}|).
\ee
and for small $w$ we have
\be
{\cal X}_w(\mathbf{b},P)\approx
\frac{1}{4\tau}
\,
\frac{ e^{-2|\mathbf{b}||\frac{P}{\sin w}|}}{\sqrt{\pi|\mathbf{b}||P\sin w|}}
\ee

 It is known  that the propagator in the conic space  can be present as a   sum of
 two terms \cite{DJ88,Hooft88}
 \be
K_\alpha(z,y;z',y';\tau)=K(z,y;z',y';\tau)+K^\prime_{\alpha}(z,y;z',y';\tau),
\label{HK2terms}
\ee
where
\be
K^\prime_{\alpha}(z,y;z',y';s)=\frac{i}{2\alpha }\int_{\gamma'} dw\,
{\mbox{ctg}\,}\left(\frac{\pi w}{\alpha}\right)\,K(z(w),y(w);z',y';s),
\label{gamma'}\ee
with a modified contour $\gamma'$ presented in Fig. \ref{Fig:contours}.

Therefore, the eikonal matrix element can be written  as
\bea
\nonumber
{\cal S}_{{\rm  eik}, \alpha} (p_1,p_2,p_3,p_4)&=&i(2\pi)^4\delta^4(p_1+p_2-p_3-p_4){\cal A}_{{\rm eik, flat}}
\\
&+&i(2\pi)^3\delta^{3}\left((p_1+p_2-p_3-p_4)_{\check{\mu}}\right){\cal M}_{{\rm eik},\alpha}, \eea
where
\be
{\cal M}_{\rm eik,\alpha} =-2i\tau \int d^2b_\perp e^{iq_\perp b_\perp}
e^{i\chi_{plane}(b_\perp)}\left[e^{\Delta\chi_\alpha(b_\perp,P)}-1\right],
\label{eicon-def}\ee
where $\chi_{\rm plane}(b_\perp)$ is given by (\ref{bbc}) for $n=1$ and
\bea
\Delta\chi _\alpha (b_\perp,P)=\frac{1}{2s} \int \frac{d^2q_\perp}{(2\pi)^2}
e^{-iq_\perp b_\perp} B_\alpha(q_\perp^2,P),
\label{eikp}
\eea
where
\be
B_\alpha(q_{\bot},P)=\frac{i}{2\alpha }\int_{\gamma'} dw\,
{\mbox{ctg}\,}\left(\frac{\pi w}{\alpha}\right){\cal B}_w(q_{\bot},P).\ee

Now if we take this correction perturbatively we get
\be
{\cal M}_{{\rm eik},\alpha} \approx-2is \int d^2b_\perp
\,\Delta\chi_\alpha(b_\perp,P)\,
e^{iq_\perp b_\perp+i\chi_{\rm plane}(b_\perp)}.
\label{eicon-def}\ee
We can analyze the correction for arbitrary angle $\alpha$ only numerically.

 \subsection{Correction to the eikonal amplitude for toy model $\alpha=\pi/N$}
 It is known, that the propagator in the conic space with $\alpha=\pi/N$ can be present as a finite  sum of
 propagators
\be
\label{HK-r-total}
K_{\pi/N}(z,z',\tau)=\sum _{n=0}^{N}{\cal K}_{n\pi/N}(z,z',\tau),
\ee
where
\be
\label{HK-r-m}
{\cal K}_{n\pi/N}(z,z',\tau)\equiv \frac{1}{4\pi \tau}
\exp\{-\frac{z^2+{z'}^2-2zz'' \cos(\frac{n\pi}{N} )}{4\tau}\}.
\ee
We can calculate the contour integral in (\ref{BA-dif1}) explicitly to get
\bea
\nonumber
{\cal S}_{\pi/N}&=&i(2\pi)^3\delta^{3}\left((p_1+p_2-p_3-p_4)_{\check{\mu}}\right)\left[
\sum \,^\prime
\,\frac{2}{\sin \frac{\pi n}{N}}\,\frac{1}{\frac{Q^2}{\sin^2 \frac{\pi n}{2N}}+
\frac{P^2}{\cos^2 \frac{\pi n}{2N}}+q_{\check{\mu}}^2+m^2}\right.\\
&+&\left.
 \delta(Q)\,
 \frac{\pi}{\sqrt{P^2+q_{\check{\mu}}^2+m^2}}+\delta(P)\,
 \frac{\pi}{2\sqrt{q_\mu^2+m^2}}\right].\label{BA-dif1-sum}
\eea
Here the prime in the sum means that we do not take into account $n=0$
and $n=N$.

If we consider $N=1$ we get just one new term as a correction to the usual Born amplitude
\bea
\label{BA-n}
{\cal S}_{\pi}&=&{\cal S}_{\rm flat}+\Delta{\cal S}_{\pi},\\
\label{flat}
{\cal S}_{\rm flat}&=&\delta^{4}(p_1+p_2-p_3-p_4)\frac{i(2\pi)^4}{2\sqrt{q_\mu^2+m^2}},\\
\label{rflat}\Delta{\cal S}_{\pi}&=&
\delta^{3}\left((p_1+p_2-p_3-p_4)_{\check{\mu}}\right)
\delta \left((p_1-p_3-p_2+p_4)_z\right)\frac{i(2\pi)^4}{2\sqrt{q^2+m^2}}.\eea
In the eikonal regime $Q\approx P$ and both terms (\ref{flat}) and (\ref{rflat})
give the same contribution and we get a doubling of the eikonal phase, see Fig.3.B.

\section{Conclusion and Discussion}
In this paper we have argued that in the TeV-gravity scenario high energy hadrons
colliding on
 the 3-brane embedded in $D=4+n$-dimensional spacetime,  with $n$ dimensions
smaller than the  hadrons size,  can be considered as  cosmic membranes. In
the 5-dimensional case
this consideration leads to the 3-dimensional effective model
of high energy collisions of hadrons. The cosmic membranes in
the 5-dimensional case are similar to cosmic
strings in the 4-dimensional world.

It is well known that the  cosmic strings give rise to remarkable
classical gravitational and quantum phenomena. In particular, the
cosmic string acts as a gravitational lens
 \cite{vilenk85}. This effect becomes manifest when two particles
 move along opposite sides of the string. Also there is a self-force
 acting on a test charged particle around the
cosmic string \cite{Linet} and a freely moving charged particle  radiates near
the
cosmic string \cite{VF-VS,Galtsov}. This is an analogue of the radiation by the
charged particle when
it suffers the Aharonov-Bohm scattering \cite{BA59} and  this radiation occurs
due to the fall down of the Huygens principle in curved spacetime.

There are also quantum effects. The presence of the cosmic string allows
effects such as particle-antiparticle pair production by a single
photon and bremsstrahlung radiation from charged particles
\cite{SkHaJa94,AuJaSk94} which are not possible in empty Minkowski
space, due to conservation of linear momentum. The conical structure
of the cosmic string spacetime is the source of momentum
non-conservation in the plane perpendicular to the string, which
permits pair production by a single photon. The
gravitational mechanism that permits pair production by a single
photon around a cosmic string has common topological features  with
the Aharonov-Bohm effect \cite{BA59}. The absence of global
momentum conservation was already stressed for gravity in 2+1
dimensions by Henneaux \cite{Marc84} and Deser \cite{deser85}. It is worth also to
mention that  the string polarizes the vacuum around it, in a way
similar to the Casimir effect between two conducting planes forming
a wedge \cite{DC79,HellKon86}. The study of quantum field theories the spasetime with conic
singularities requires a regularization \cite{cone-zeta}. Among possible regularizations
the zeta-function regularization
is more convenient \cite{zeta}.

Our specifics is that not all process mentioned above can be realized for
particles attached to the 3-brane. In particular, to see the lens effect we have to deal with
the motion of  particles in the 2-plane that is  perpendicular  to the hadron membrane.
 But only gravitons can move in this plane in any direction.
 However one can estimate  the self-force effect.

The same concerns also the quantum effects. From one side, only the graviton
 can propagate in the  2-plane  perpendicular to the hadron membrane and
 feel   the deficit angle.
From other side, the above mentioned quantum processes are available
for other particles if their have not to abandon the 3-brane to participate in the processes.

In this paper we have estimated corrections to the eikonal scattering amplitude
due to the hadron membrane.

Similar to the case of cosmic string \cite{SkHaJa94},
one can also estimate the decay of a
 light ultra-relativistic particle on two heavy particles with mass $M$.
 For large longitudinal momentum of the light particle,
$k_z>>2M\delta^{-1}$, the cross-section   does not depend on $k_z$ and is defied
only by the coupling
$g$ of these 3 particles and heavy mass
\be
\sigma _{1\,{\rm light}\to 2 \,{\rm heavy}}\approx \frac{g^2}{M^3}
\ee
To realize the condition  $k_z>>2M\delta$ it is enough to take $k_z\sim 1 TeV$ and $M$
of the order of the few $MeV$'s.

Other processes we are going to estimate in the separate work \cite{IA-AB}.
\section*{Acknowledgments}

It my  pleasure to present this paper to Proceedings of Workshop:
"Cosmology, the Quantum Vacuum, and Zeta Functions" dedicated to Emilio Elizalde's
sixtieth birthday.

I am  grateful to A.V. Radyushkin, I.V.Volovich and A.A.Bagrov for useful discussions.
This work  is supported in part by RFBR grants 08-01-00798 and 09-01-12179-ofi$_m$
and  by state contract of Russian Federal
Agency for Science and Innovations 02.740.11.5057.


{\small
\begin{thebibliography}{20}
\bibitem {ADD}N.~Arkani-Hamed, S.~Dimopoulos and G.~R.~Dvali,
Phys.\ Lett.\ B \textbf{429} (1998) 263, hep-ph/9803315;
I.~Antoniadis, N.~Arkani-Hamed, S.~Dimopoulos and G.~R.~Dvali,
Phys.\ Lett.\  B {\bf 436} (1998) 257,
hep-ph/9804398;
\bibitem {factory} S.~Dimopoulos and G.~Landsberg,
Phys.\ Rev.\ Lett.\ \textbf{87} (2001) 161602, hep-ph/0106295;
S.~B.~Giddings and S.~Thomas,
\ Phys.\ Rev.\ D \textbf{65} (2002) 056010,
hep-ph/0106219.

\bibitem{GRW98}
G.~F.~Giudice, R.~Rattazzi and J.~D.~Wells,
Nucl.\ Phys.\ B {\bf 544} (1999) 3, hep-ph/9811291.
\bibitem {GRW01}G.~F.~Giudice, R.~Rattazzi and J.~D.~Wells,
\ Nucl.\ Phys.\ B \textbf{630} (2002) 293, hep-ph/0112161.



\bibitem {Lonnblad} L.~Lonnblad and M.~Sjodahl,
JHEP \textbf{0610} (2006) 088, hep-ph/0608210.
\bibitem{Boelaert:2009jm}
  N.~Boelaert and T.~Akesson,
  Eur.\ Phys.\ J.\  C {\bf 66} (2010) 343, arXiv:0905.3961.

\bibitem{Sessolo:2008ex}
  E.~M.~Sessolo and D.~W.~McKay,
  Phys.\ Lett.\  B {\bf 668} (2008) 396, arXiv:0803.3724.


\bibitem{SRychkov}
  P.~Lodone and S.~Rychkov,
  JHEP {\bf 0912} (2009) 036, arXiv:0909.3519.
  \bibitem {BHinCR}J.~L.~Feng and A.~D.~Shapere,
Phys.\ Rev.\ Lett.\ \textbf{88}, 021303 (2002),
hep-ph/0109106.
\newline L.~Anchordoqui and H.~Goldberg,
Phys.\ Rev.\ D \textbf{65}, 047502 (2002), hep-ph/0109242.


\bibitem{EMR01}
R.~Emparan,
Phys.\ Rev.\ D {\bf 64} (2001) 024025;
R.~Emparan, M.~Masip and R.~Rattazzi, Phys.Rev. D65 (2002) 064023,
hep-ph/0109287.
\bibitem {BF99}T.~Banks and W.~Fischler,
{\it "A model for high energy scattering in quantum gravity"}, hep-th/9906038.
\bibitem{IA99}  I.Ya. Aref'eva,
Part.Nucl.31 (2000) 169, hep-th/9910269.

\bibitem {EG02}D.~M.~Eardley and S.~B.~Giddings,
\ Phys.\ Rev.\ D \textbf{66}, 044011 (2002),
gr-qc/0201034.


\bibitem {GR04}S.~B.~Giddings and V.~S.~Rychkov,
Phys.\ Rev.\ D \textbf{70}, 104026 (2004),
hep-th/0409131;
\textit{ibid.} \textbf{436}, 257 (1998), hep-ph/9804398
\bibitem{IA09}  I.Ya. Aref'eva, {\it Catalysis of Black Holes/Wormholes Formation  in High Energy
Collisions},  arXiv:0912.5481
\bibitem{AV-TM}
  I.~Y.~Aref'eva and I.~V.~Volovich,
  Int.\ J.\ Geom.\ Meth.\ Mod.\ Phys.\  {\bf 05} (2008) 641, arXiv:0710.2696.
  \bibitem{MMT}
  A.~Mironov, A.~Morozov and T.~N.~Tomaras,
 {\it If LHC is a Mini-Time-Machines Factory, Can We Notice?},
arXiv:0710.3395.
  \bibitem{NS}
  P.~Nicolini and E.~Spallucci,
  {\it Noncommutative geometry inspired wormholes and dirty black holes},
  arXiv:0902.4654.

\bibitem {tHooft}G.~'t Hooft,
\ Phys.\ Lett.\ B \textbf{198},
61 (1987).
\bibitem {ACV87} D.~Amati, M.~Ciafaloni and G.~Veneziano,
Phys.\ Lett.\ B \textbf{197}, 81 (1987).


\bibitem {Verlindes}H.~L.~Verlinde and E.~P.~Verlinde,
Nucl.\ Phys.\ B \textbf{371}, 246 (1992), hep-th/9110017.


\bibitem {Kabat92} D.~N.~Kabat and M.~Ortiz,
Nucl.\ Phys.\ B \textbf{388}, 570 (1992), hep-th/9203082.

\bibitem {Amati:1987uf}D.~Amati, M.~Ciafaloni and G.~Veneziano,
Int.\ J.\ Mod.\ Phys.\ A \textbf{3}, 1615 (1988),
\ Nucl.\ Phys.\ B \textbf{347}, 550 (1990).


\bibitem {Amati:1992zb} D.~Amati, M.~Ciafaloni and G.~Veneziano,
Phys.\ Lett.\ B \textbf{289}, 87 (1992),
JHEP \textbf{0802}, 049 (2008), arXiv:0712.1209.

\bibitem{areva}
I.~Ya.~Aref'eva, K.~S.~Viswanathan and I.~V.~Volovich,
Nucl.\ Phys.\ B {\bf 452}, 346 (1995).
\bibitem {Ciafaloni:2008dg} M.~Ciafaloni and D.~Colferai,
JHEP \textbf{0811}, 047 (2008),
arXiv:0807.2117.





 \bibitem{RS2}
  L.~Randall and R.~Sundrum,
  Phys.\ Rev.\ Lett.\  {\bf 83}, 4690 (1999), hep-th/9906064.
 \bibitem{DGP}
  G.~R.~Dvali, G.~Gabadadze and M.~Porrati,
  Phys.\ Lett.\  B {\bf 485}, 208 (2000), hep-th/0005016.

  \bibitem{IA}  I.Ya. Aref'eva, {\it Phys.Lett.},  {\bf B 325} \ (1994) 171, hep-th/9311115.
\bibitem{LL}  L.N. Lipatov, JETP Lett. 59 (1994) 596;
R.~Kirschner, L.~N.~Lipatov and L.~Szymanowski,
  Nucl.\ Phys.\  B {\bf 425}, 579 (1994), hep-th/9402010.

  \bibitem{vilenk85} A. Vilenkin,
  Phys. Rep. {\bf 121}, 263 (1985).

\bibitem{0809.0005}   O.~Pujolas,
  JHEP {\bf 0812}, 057 (2008), arXiv:0809.0005.

 \bibitem{IA-AB}  I.Ya. Aref'eva and A.A.Bagrov, in preparation.
\bibitem{Eikonal}
  M.~Levy and J.~Sucher,
  Phys.\ Rev.\  {\bf 186} (1969) 1656.


 \bibitem{Eikonal-dubna} B.M.Barbashev, S.P.Kuleshev, V.A.Marveev and A.N.Sissakian,
 Theor.Math.Phys., 3 (1970) 342.
\bibitem{DJH84}  S.~Deser and R.~Jackiw, G. 't Hooft,
Ann. of Phys., NY., 152 (1984) 220.
\bibitem{Dowker77}
J.S. Dowker,
J.Phys. {\bf A10} (1977) 115
\bibitem{DJ88}  S.~Deser and R.~Jackiw,
  Commun.\ Math.\ Phys.\   118 (1988) 495.
\bibitem{Hooft88}  G. 't Hooft,
Comm.\ Math.\ Phys, \ 117 (1988) 685.



\bibitem{Linet}  B. Linet,
Phys.Rev, 33 (1986) 1833.
\bibitem{VF-VS}  V.~P.~Frolov,  E.~M.~Serebryanyi and V.~D.~Skarzhinsky,
  ``{\it On the electrodynamical effects in the gravitational field of a cosmic
  string},''
in "Proceedings of quantum gravity conference",  Moscow, (1987) 830.

\bibitem{Galtsov} A.N. Aliev, D.V. Galtsov,
Zh.Eksp.Teor.Fiz. 96 (1989) 3.
\bibitem{BA59} Y. Aharonov and D. Bohm, Phys. Rev.,
119 (1959) 485.


\bibitem{SkHaJa94}V.D. Skarzhinsky, D.D. Harari, and U. Jasper,
Phys. Rev. D 49 (1994)  755.
\bibitem{AuJaSk94}
  J.~Audretsch, U.~Jasper and V.~D.~Skarzhinsky,
  Phys.\ Rev.\  D 49 (1994) 6576

\bibitem{Marc84} M. Henneaux,
Phys.Rev.D 29 (1984) 2766.
\bibitem{Audretsch:1994ua}
  J.~Audretsch, U.~Jasper and V.~D.~Skarzhinsky,
  Phys.\ Rev.\  D 49 (1994)  6576.

\bibitem{deser85}
S. Deser,
 Class.Quant.Gr., 2 (1985) 489.

\bibitem{DC79} D. Deutsch and P. Candelas,
Phys. Rev. D 20 (1979) 3063.



\bibitem{HellKon86} T.H. Helliwell and D.A. Konkowski,
Phys. Rev. D 34 (1986) 1918.
\bibitem{cone-zeta}  G.Cognola, K.Kirsten,  L. Vanzo,
Phys. Rev. D 49 (1994) 1029.

\bibitem{zeta} E. Elizalde, S. D. Odintsov, A. Romeo, A. A. Bytsenko, and S. Zerbini,
{\it Zeta regularization
techniques with applications}, World Scientific Publishing Co. Inc., River Edge, NJ, 1994.


\end{thebibliography}
}
\end{document}